# Anomalously augmented charge transport capabilities of bio–mimetically transformed collagen intercalated nano–graphene based biocolloids

Purbarun Dhar[1, #], Soumya Bhattacharya[2], Suprabha Nayar[2] and Sarit K. Das[1,]*

[1] Department of Mechanical Engineering, Indian Institute of Technology, Madras, Chennai – 600 036

[2] Biomaterials Group, Materials Science and Technology Division, National Metallurgical Laboratory (CSIR), Jamshedpur – 831 007

[#] Electronic mail: pdhar1990@gmail.com

* Corresponding Author: Electronic mail: skdas@iitm.ac.in

Phone: +91-44-2257 4655

Fax: +91-44-2257 4650

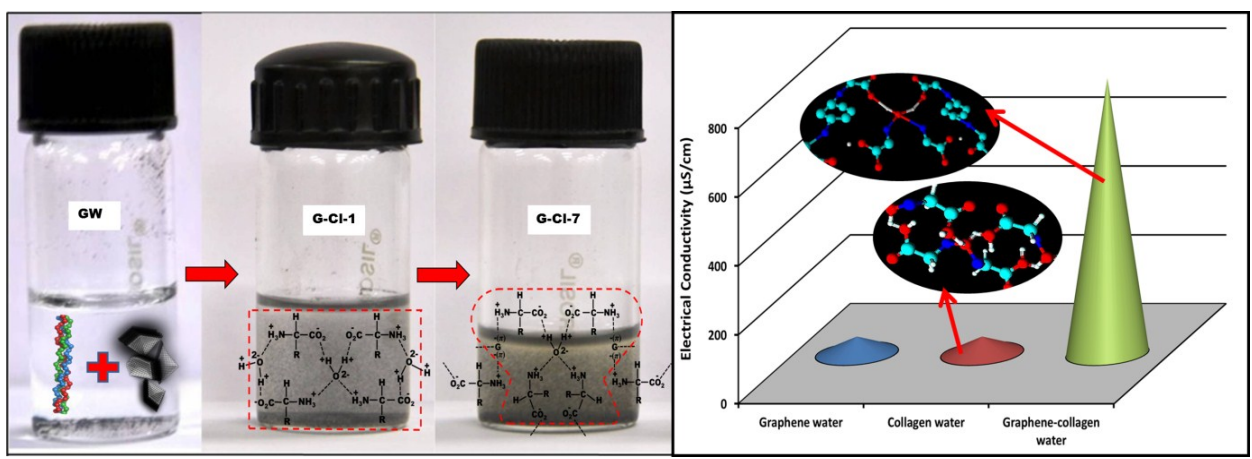

**Graphical Abstract**

# Abstract

Collagen micro-fibrils bio–mimetically intercalate graphitic structures in aqueous media to form graphene nano-platelets collagen complex (G-Cl). Synthesized G-Cl based stable, aqueous bio-nanocolloids exhibit anomalously augmented charge transportation capabilities over simple collagen or graphene based colloids. The concentration tunable electrical transport properties of synthesized aqueous G-Cl bio-nanocolloids has been experimentally observed, theoretically analyzed and mathematically modeled. A comprehensive approach to mathematically predict the electrical transport properties of simple graphene and collagen based colloids has been presented. A theoretical formulation to explain the augmented transport characteristics of the G-Cl bio-nanocolloids based on the physico-chemical interactions among the two entities, as revealed from extensive characterizations of the G-Cl bio-complex, has also been proposed. Physical interactions between the zwitterionic amino acid molecules within the collagen triple helix with the polar water molecules and the delocalized $\pi$ electrons of graphene and subsequent formation of partially charged entities has been found to be the crux mechanism behind the augmented transport phenomena. The analysis has been observed to accurately predict the degree of enhancement in transport of the concentration tunable composite colloids over the base colloids. The electrically active G-Cl bio-nanocolloids with concentration tunability promises find dual utility in novel gel bio–electrophoresis based protein separation techniques and advanced surface charge modulated drug delivery using biocolloids.

# Keywords

Nano Graphene, Collagen, Graphene-collagen bio composite, nanofluids, electrical conductivity, drug delivery.

# Introduction

Nano-allotropes of carbon such as carbon nanotubes, fullerenes, nano-carbon and graphenes have stormed the academic community in the recent decade owing to their interesting and unique physical and chemical structure, functionalities and properties [1, 2, 3, 4]. On the applied front, graphene based nano–electronics [5, 6] and opto–electronic photonics devices [7, 8] have shown tremendous promise as next the generation smart devices. However, in more recent times, graphene has found its way into the realms of nano–biology research in view of the tunable properties that can be modulated to alter or harness variant biological and/or physiological processes and operations [9, 10, 11]. Several research initiatives to better comprehend the associativity and interactivity of graphene sheets with biological entities such as cancerous tissues [12, 13, 14] amino acids and DNA [15, 16, 17] and the associated exploitable characteristics have been explored towards the end of past decade. Researchers have provided concrete evidence that if harnessed and bio–functionalized in accordance to standard protocols, graphene structures can be utilized as vector agents for future nanomedicine. Owing to the high dispersibility of functionalized (in general carboxylated or sulphonated) graphene nano-platelets in aqueous systems [18], research in the area of graphene based colloids has also seen immense popularity, since administering colloidal systems into biological entities offers higher ease of modulation and manipulation. Enhanced thermo-physical properties [19, 20, 21] of such colloids have been reported experimentally and predicted through analytical models [22, 23]. The charge transport characteristics of nano-graphene based colloids have also been reported [24], however, the nature and mechanism of augmented transport is not comprehensive and lacks any mathematical formulation or treatment. Furthermore, the degree of augmentation in conjunction with bio–entities is also an area where light needs to be shed for proper design of future graphenic biocolloids. The first section of the present article provides a mechanistic outlook and mathematical prediction of the enhanced transport phenomena in basic graphenic colloids, which is essential towards understanding the phenomena of its bio interactions and subsequent applications.

Collagen is the major building constituent of biological tissues and its interactions with nanostructures and materials has been an issue of academic interest in recent times owing to the rising promise of nanomedical sciences. Since collagen is a major constituent of tissues that build up the human body, biocompatibility of a nanostructure with collagen is of prime importance for it to qualify as a future nanomedicine vector. Few investigations on the mechanical properties and characteristics of collagen-graphene structures have been reported in very recent times [25],[26]. Sub-micron scale graphitic flakes (~ 10 micron flake size) have been proven to be bio-mimetically exfoliated to nano-graphene (2–5 layers, flake size distribution peak at ~ 300 microns) by collagen microfibrils in a novel methodology reported previously by the present authors [27]. It was observed that a graphite-collagen aqueous colloid spontaneously or bio-mimetically transforms into a nano-graphene-collagen aqueous colloid when left undisturbed for few days. Detailed characterization provided evidence of graphene-collagen intercalation and subsequent exfoliation of graphite to nano-graphene structures. Furthermore, observations revealed that collagen fibrillated nano-graphene based aqueous biocolloids exhibit anomalously enhanced charge transport characteristics in comparison to simplistic graphene or collagen based colloids. Such enhanced transport properties, if comprehended mechanistically, could be harnessed in variant charge transport based bio–processes, viz. gel based electrophoretic sampling or separation of bio–molecules, DNA manipulation through charge identification, cancer cell targeting and drug delivery through charged drug molecules etc. and therefore requires a comprehensive theory. The second section of the article provides a theoretical understanding and identifies the mechanism behind such anomalous enhancements.

# Materials and Methodologies

## Synthesis of G-Cl biocolloids

The G–Cl complex based biocolloids are synthesized in accordance to protocol reported in results published by the present authors[27]. The entire process is biomimetic since it requires no external stimuli in order to achieve the transformation of graphite to nano–graphene and subsequent formation of G-Cl complex. In précis, high purity graphite powder (mesh size ~ 75 microns, Alfa Aesar, USA) was added to a aqueous colloidal system of Type 1 collagen derived from the Achilles tendon of bovines (~ 300 nm fibril length and ~ 1.5 nm fibril diameter, Sigma Aldrich, USA), such that graphite : collagen weight ratio was 1:1. The samples were allowed to stand undisturbed and characterized 1 day and 7 days later. Five different samples with the G-Cl concentration ranging from 0.01 wt. % to 0.05 wt. % were prepared. The transformation of graphite to graphene by intercalation based exfoliation by collagen strands has been observed to be complete within a week, with the samples characterized after day 1 showing partial transformation and those after 7 days showing no further change than that of day 7. The transformation can be observed visually as illustrated in Fig. 1. (a)–(c). The samples were centrifuged to collect the G-Cl bio–complex for characterizations.

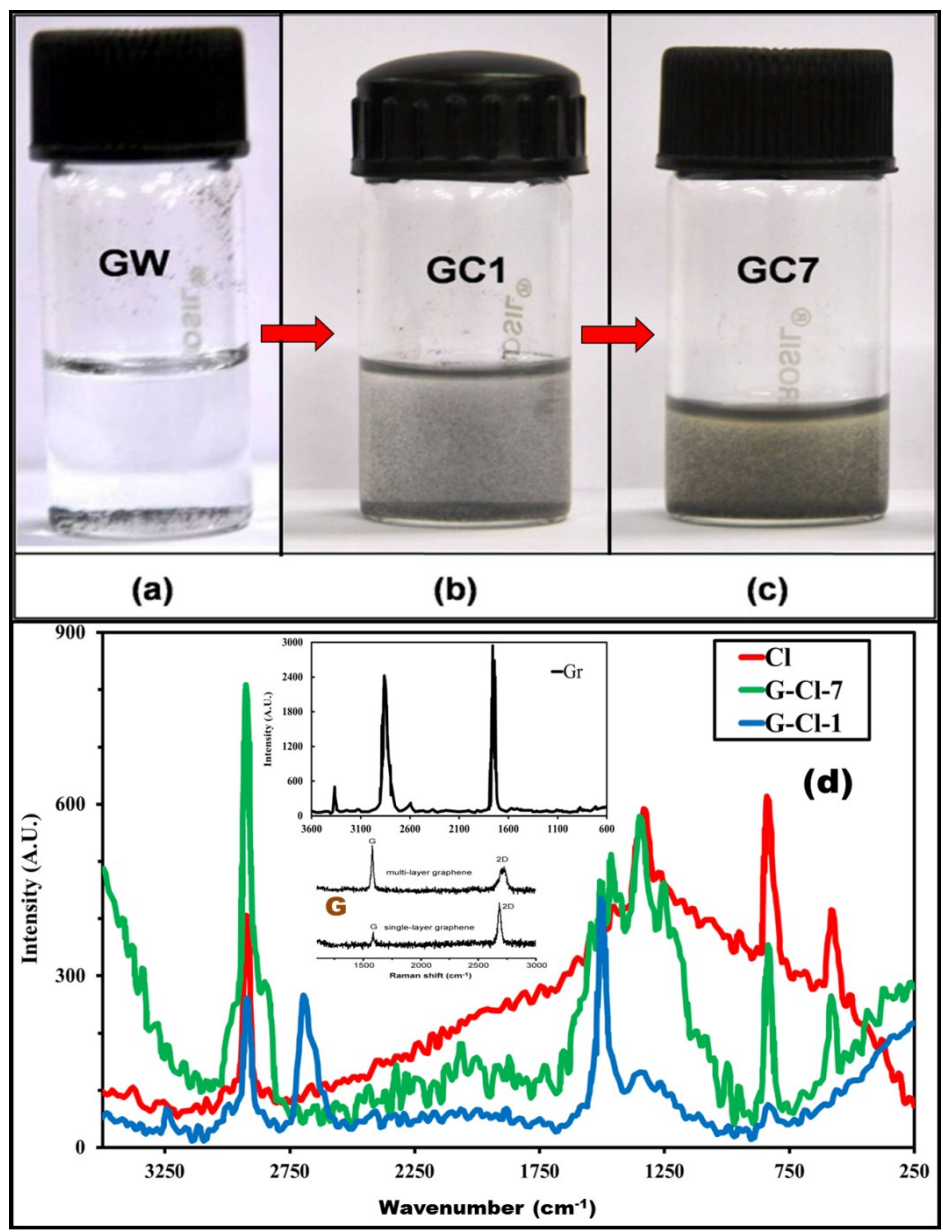

**FIGURE 1**: Spontaneous bio-mimetic transformation of graphite to nano-graphene by collagen micro-fibrils. **(a)** Initial graphite-collagen aqueous (GW) colloid. The colloid exhibits low stability. **(b)** The same sample after day 1 (G-Cl-1). The exfoliation of graphite to graphene begins and a stable collagen-graphene colloid ensues, as observed from the gradual darkening of the colloid. **(c)** The solution after day 7 (G-Cl-7). The exfoliation process is complete and a very stable collagen-graphene colloid is formed. The colloid is centrifuged at above 10,000 r.p.m to separate the graphene-collagen complex. It can be observed from the figures that the hydrophilic nature of the complex increases and leads to highly stable biocolloids. Snapshots of different samples have been utilized and hence the non–equity in volume is observed. **(d)** Surface Enhanced Raman Spectra of Collagen (Cl), Graphene-Collagen complex after day 1 (G-Cl-1) and after day 7 (G-Cl-7). The SERS for graphite (Gr) and single and multi-layer graphene (G) have been provided inset.

**Characterization of G-Cl biocolloids**

The G-Cl biocomposite samples under observation have been characterized in order to understand the physico–chemical interactions which are essential in order to comprehend and model the augmented charge transport phenomena. Formation of graphene from graphitic structures and by intercalation based exfoliation by the collagen fibrils, and subsequent transformation to a stable G-Cl complex has been revealed from Surface Enhanced Raman Spectra (SERS). The SERS, illustrated in Fig. 1(d), has been obtained by a dispersive Raman spectrometer (Thermo Fisher Scientific, USA) employing a 532 nm Nd-YAG laser source employing a droplet of the G-CL biocolloid on a polished copper substrate as the target. Distinct graphitic (Gr) and collagen based colloids have been used as the controls. The Gr exhibits distinct G and 2D bands at 1592 and 2728 cm$^{-1}$ respectively, and low defect Gr can be inferred from the absence of D band. However, the D peak was observed to appear in G-Cl-1 at ~ 1360 cm$^{-1}$ and enhanced in G-Cl-7, indicating augmented edge chirality [27], implying that the graphitic sheet ordering has been disrupted by the presence of collagen microfibrils. This proves evidence towards ripping of the low disordered Gr sheets by collagen. The intercalation by Cl led to repulsive action between the delocalized pi electron cloud in graphite flakes and the negative poles of the constituent zwitterionic amino acid molecules, thereby leading to transient weakening of the stacking van der Waals forces between the inter planar $p_z$-$p_z$ orbitals. The G band showed no change in the transformed composites, however, the 2D band in Gr was observed to decrease in intensity and broaden in G-Cl-1 and was nearly absent after day 7. This is representative of the fact that the stacking order is lost and the planar layer separation has increased from the constant of ~ 0.345 nm, which essentially is indicative of the fact that graphitic stacks have been distorted and separated and collagen fibers have intercalated within the layers of the separated stacks. The intercalative action leads to separation of graphitic stacks to graphenic structures complex with the collagen. The ripping action also leads to formation of nano graphene platelets from mesoscale graphite, which has been discussed in subsequent sections. The characteristics revealed by the SERS is very similar to that reported by the authors previously [27], hinting at efficient usage of protocol. The major peaks obtained from the SERS all reveal the presence of graphenic structures in the colloidal system. The 2D band system in G-Cl-1 comprises of many distinctly identifiable overlapping peaks which speaks of graphenic structures of the order of ~ 5 or less layers[27]. The temporal increase of the $I_D/I_G$ ratio from 0 in Gr to 0.27 in G-Cl-1 and 0.99 in G-Cl-7 provides evidence of rearrangements and reshuffling based transformation of the G-Cl complex from nanocrystalline to amorphous collagen complexed graphenic structures.

Further evidence for the transformation of graphitic structures to graphenes and its formation of physico–chemically bonded complex with collagen microfibrils has been obtained via X-ray Diffraction (XRD), Fluorescence (FLS) and Photoluminescence (PLS) Spectroscopy. XRD was performed utilizing a Diffractometer (Billerica, USA) at 40 kV employing Cu–kα radiation of wavelength 1.5418 Å with 0.02 °/step and scan rate of 5 s/step. The XRD spectra have been illustrated in Fig. 2(a). Cl expectedly showed the amorphous hump, uncharacterized by the presence of any peaks. G-Cl-1 and G-Cl-7 showed sharp peaks at ~ 26.598 ° which corresponds to the (002) Bragg plane of Gr structures. As evident from Fig. 2(a), the peak intensity is low in G-Cl-1 and increases drastically in G-Cl-7. In powder diffraction method, single layer graphene is not expected to exhibit any peak and the peak intensity augments proportionally to the number of layers[27]. Since the concentration of graphite and collagen in G-Cl-7 and G-Cl-1 are equal, the triple intensity scale of the former compared to the latter reveals increase in average crystallite size with time and possible exfoliation of graphite followed by self-assembly of the graphenic structures under the influence of collagen intercalation. The two phenomena combined hints towards the fact that collagen intercalated within the graphitic stacks and ripped them apart into highly distorted (as revealed from the increased crystallite size) nano-graphenic structures. With time, these reassembled into aggregates (derived from the fact that the G-Cl-7 has an overall higher intensity count, which indicates enhanced crystalline nature, hinting at assembled pseudo-amorphous structures) due to the enhanced interfacial charge developed at the water-graphene-zwitterionic collagen complex, caused by the increases electrostatic interactions between the amino acid molecules and the exposed pi electrons of graphene structures due to intercalation and exfoliation (discussed in detail in subsequent sections). The aggregates formed the stable G-Cl complex which is responsible for the enhanced peak intensity observed and subsequent formation of augmented interfacial charge in presence of a polar dispersion medium.

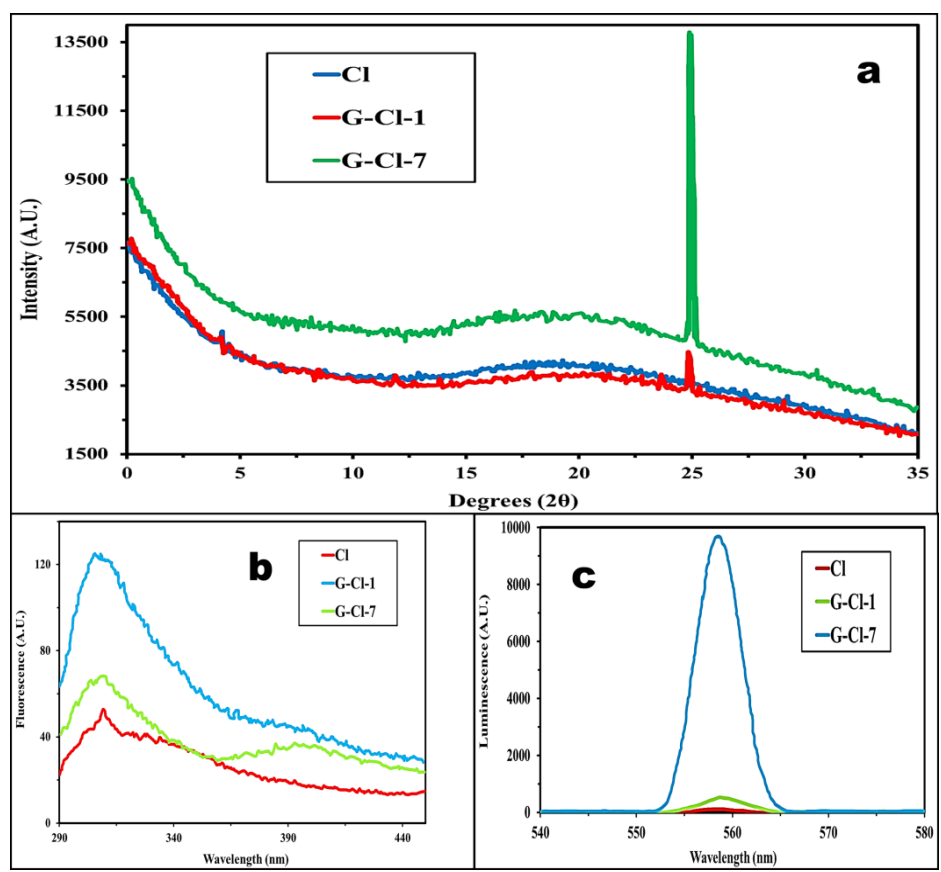

**FIGURE 2**: Characterization spectra for Cl, G-Cl-1 and G-Cl-7 systems **(a)** XRD spectra **(b)** Fluorescence spectra **(c)** Photoluminescence spectra.

Spatial conformation of protein molecules can be identified though FLS and PLS of the concerned material. FLS and PLS has been performed on the G-Cl bio–complex in order to understand the structural modifications that the Cl strands undergo when associated with graphene structures. The FL spectrum has been illustrated in Fig. 2(b) and as can be observed, the spectrum has highest recorded intensity count for G-Cl-1 followed by G-Cl-7 and least for Cl. The PL spectrum has been illustrated in Fig. 2(c) and it has been observed that G-Cl-7 exhibits peak intensity much higher than its G-Cl-1 or Cl counterparts. The revelations from the two spectra can be analyzed simultaneously to understand G-Cl temporal interactivity, which in turn leads to augmented charge transport. FL spectrum is obtained due to the aromatic structures present in a system. The increased FL in G-Cl-1 is possibly due to the presence of the aromatic structures in Gr, which are ruptured apart with time as the Cl intercalates into Gr to form graphenic systems, as observed from the sharp decrease in FL in G-Cl-7. G-Cl-7 also develops a distinct bulged peak at ~ 400 nm, which is not present in either Cl or G-Cl-1. This is postulated to appear due to interactivities between the amino acid molecules in Cl and graphene, wherein, the largely delocalized pi electrons in the G-Cl system leads to charge interactions between graphene and Cl. This in essence leads to higher charge transport capabilities in the colloidal state and has been explained in later segments. Gr to graphenic transformations are also supported by the PL spectrum. PL emission is in general higher for open structures, which indicates that the openness in G-Cl-7 is much higher that G-Cl-1, which proves that over time, Cl has been successful in cleaving the Gr sheets along horizontal planes to create nanoscale sheets, thereby increasing number of exposed edges. Furthermore, single layer pristine graphenes are not expected to exhibit PL, since it is derived from localized electron–hole interactions in $sp^2$ systems[27]. As the number of layers stack towards Gr, the effective surface area available for PL emissions to be detectable decrease drastically as the emissions within the internal layers are unable to escape the stacked system. Thereby, multilayered graphenes, ranging from bilayer to penta-layered systems are expected to emit high intensity PL spectrum, which is the case of G-Cl-7. Evidently, the analysis of FLS and PLS shows that with time, Cl has successfully exfoliated and ruptured graphitic flakes to graphenic flakes and simultaneously intercalated via electrostatic interactions to form the G-Cl complex.

Evidence of formation of graphenic structures due to intercalation by Cl can also be qualitatively obtained from confocal and electron microscopy. As reported by the present authors[27], the thickness contours of the G-Cl-7 obtained through Transmission Electron Microscopy (TEM) revealed sheets of average thickness of 5. In the present paper, TEM images of the composite have been illustrated (Fig. 3(b)) which reveals the morphological features of the G–Cl system; which are basically surface contours signifying stacking faults with the intercalation of collagen fibrils within the graphene layers. A region of Fig. 3(b) has been studied exclusively so as to determine the physical thickness of the formed G-Cl system. Based on the contour imaging in Fig. 3(a), the graphenic structures formed can be deduced to be 3-5 layers thick on an average. The formation of graphene can also be confirmed from the indexed SAED pattern corresponding to the image shown in Fig. 3b; the image shows the first and second order Laue zones. The hexagonal symmetry of the pattern is typical of Gr, however, the intensity of the spots corresponding to {1120} planes appears brighter than the {1010} planes. This observation has been well reported for few layered graphene[7,27].

Confocal microscopy images also reveal interesting information about the microstructure and interactions of graphenic structures with Cl. In a colloidal microenvironment, the dangling or exposed ends of the zwitterionic amino acid molecules in the Cl fibrils undergo charge exchange based stabilization with the dangling edge atoms of the graphenic structures, thereby leading to formation of physical attachments. This has been established through confocal microscopy[27] and has been found to have immense contributions towards the enhanced charge transport of the colloidal state. DIC mode confocal images of clearly visible isolated G-Cl flakes have been illustrated in Fig. 3(c1) and (c3). DIC allows distinguishing the components based on the refractive indices of the involved materials. Gr, Cl and graphene has been reported to possess different magnitudes of refractive indices[27], and thereby the confocal images Fig. 3(c1) and (c3) exhibit Cl fibrils, Gr and graphenic regions as white, greyish and blackened regions respectively. The distinction can be better understood in the fluorescence mode illustrated in Fig. 3(c2) and (c4) wherein graphenic regions exhibit darker zones compared to the brighter field graphitic and Cl zones. The confocal and TEM images, though at different length scales of imaging, exhibit clear evidence of physical intercalation of Cl with the graphite system that ultimately leads to formation of graphenic systems, evident from the presence of auto fluorescence signals illustrated in Fig. 3 (c5) and (c6) of G-Cl-1 due to defect induced hole-electron pair interactions in the wrinkled and rippled graphene sheets.

The stability of the G-Cl colloidal system is obtained from Zeta potential measurements, as obtained using Dynamic Light Scattering (DLS) measurements and has been illustrated in Fig. 3(d). The aqueous graphene colloids used as controls contained dispersed non-functionalized graphene. It can be observed that while non-functionalized graphene colloids, even in the highly dilute regime, exhibit very low stability, fairly concentrated G-Cl colloids exhibit stability ranging outside the conventional unstable domain of zeta potential. Charge conductance of the synthesized G-Cl biocolloids was measured utilizing a Cyberscan CON 11 electrical conductivity probe in accordance to protocol described in details in published results by the authors [28]. The charge transport phenomena of the biocolloids have been characterized for different G-Cl concentrations at different temperatures. Finally, the crux of the exfoliation mechanism maybe summarized. The chemical structure of Collagen (Cl) and all other exfoliants used so far have a striking similarity with respect to the hydrocarbon chain length and aromatic rings which are instrumental in providing the required shear for exfoliation. Cl fibrils consist of triple helical polypeptides with dimensions of ~ 300 × 1.5 nm, individual helices arranged in parallel arrays with a 67-nm axial shift with unwound helices at the two ends of 0.5nm, which is a proximal match to the graphitic (Gr) inter-layer spacing (0.334nm). The periodic gaps allow Cl to attack Gr from many sides simultaneously wherein the rupture of inter-strand H-bonds post acylation in Cl imparts particularly large rupture forces. Zigzag edges of Gr are present due to absence of C-C bond as a particular graphitic layer is out of plane with the one below. These sites are very often the site for attack by foreign atoms. Covalent functionalization at the graphitic edges with bulky groups is expected to induce strain that result in exfoliation. When such a bulky group is a protein, major dislocation of the electron cloud occurs which allows Cl helices to force its way through the graphitic stacks and exfoliate them.

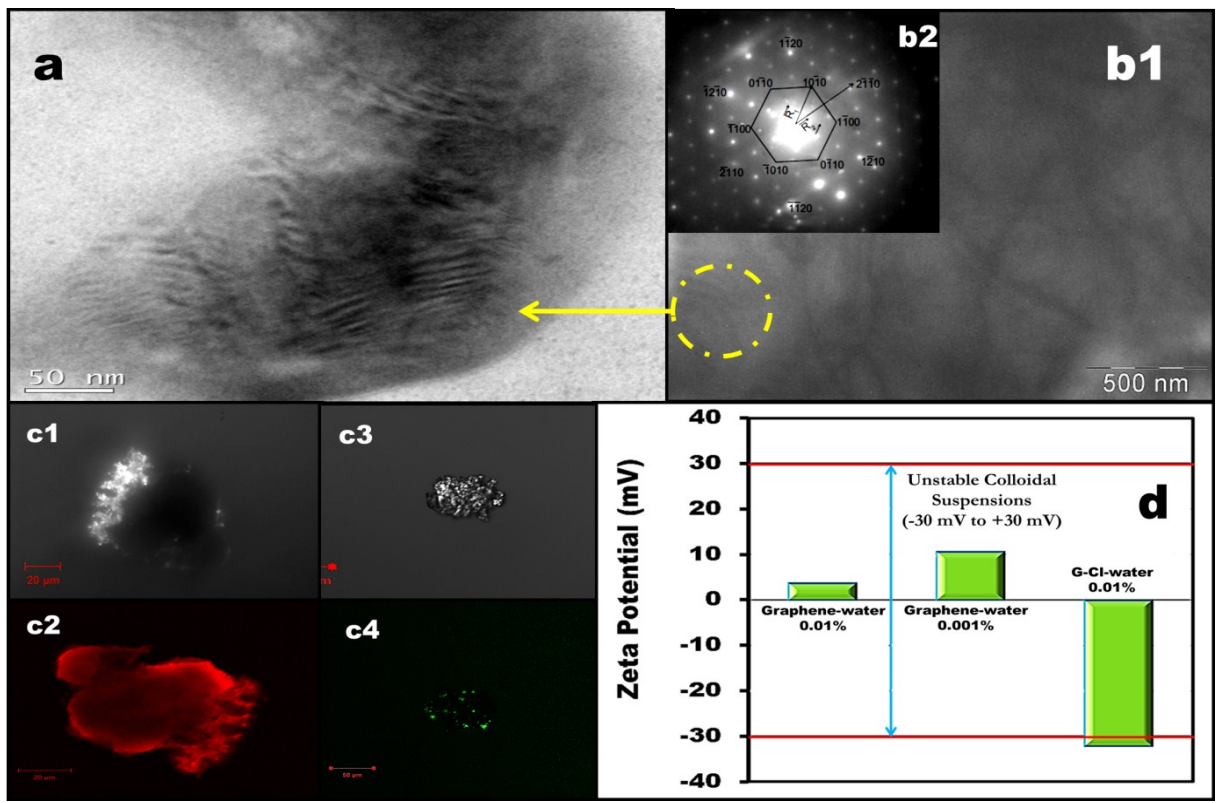

**FIGURE 3**: Imaging of G-Cl systems to visualize physical intercalation which essentially is the sole responsible mechanism for exfoliation of Gr stacks to graphenic systems. **(a)** Thickness contours obtained via TEM reveal graphenic structures few layers thick as a result of the biomimetic exfoliation by the Cl fibrils. **(b1)** Intercalation network across and within the G-Cl visible from high contrast TEM characterization, thereby revealing that the proposed exfoliation actually occurs and is the crux of the exfoliation mechanism. **(b2)** Selected Area Electron Diffraction (SAED) pattern of the graphene samples. **(c)** 1 exhibits DIC mode confocal images of an isolated flake of the G-Cl complex whereas 2 exhibit the fluorescent mode images for the same. Together, the images reveal that physical attachment to Cl to graphene, driven by fibril and flake charge stabilization, leads to formation of the stable complex, which in turn endorses stability to the colloid. Difference in optical properties also enables distinguishing between the Cl, Gr and graphene regions. 3 illustrate a singular G-Cl-1 flake and 4 represent its auto fluorescence, which reveals presence of few layer graphene spots formed in the graphite flake. **(d)** Zeta potential of graphene and G-Cl based aqueous colloids as an indicator for colloidal stability and interfacial electrical properties.

## Results and Discussions

### Analytical Formulation for Graphene nanocolloids

In an attempt to understand the behavior of charge transport in graphene–collagen colloids, it poses a necessity to comprehend charge transport in graphene colloids. Reported experimental observations[24] imply an increased charge transport ability of graphene nanocolloids over the polar dispersion fluid. To understand the physics behind such enhanced transport, a mathematical model is required to predict the augmented transport from system parameters. The population of graphene nanoplatelets within the colloid can be theorized to transport charge by two dominant mechanisms which are independent in origin yet simultaneous in existence[28]. The net electrical conductivity of the colloid is a manifestation of the twin modes of transport and their associated interactivities. The first dominant mode consists of electrophoretic drift of the nanoplatelets under the influence of the external field. It is noteworthy that appreciable electrophoretic drift is possible only for platelets whose sheet lengths are lesser than or equal to the critical sheet size[23]. For sheets sizes larger than the critical magnitude, the contribution to charge transport is significantly low due to very low streaming velocities under the influence of electric field. This is one of the reasons for lower electrical conductivities of wet synthesized graphene nanocolloids as compared to metal and metal oxide nanocolloids[28] despite the higher stability. The second dominant mode is borne from the polarization of the nanoplatelets in presence of the electric field, leading to enhanced electrostatic interactions among neighboring sheets[28]. The polarized sheets, in conjunction with the existent drift, exhibit increased transport velocities due to increased inter sheet interactions, thereby enhancing the transport phenomenon. A qualitative description of the phenomena has been illustrated in Figure 4(a).

The analytical formulation, in accordance to the bi-modal approach[28], encompasses the major and dominant modes of enhanced charge transport over the dispersing medium. It is noteworthy that in the present case, the Brownian randomness plays a minimal role in the transport process since graphene flakes; owing to their sub-micron sizes, are in general weak Brownian entities and therefore responds weakly to thermal fluctuations. Since charge transport phenomenon in colloidal state is borne of development of interfacial surface charges, it deems a necessity to incorporate the structural surface area in the analysis. For the analysis, the effective surface area on which interfacial charge is developed should be determined, since the bi-modal approach has been derived for spherical nanostructures. Based on equivalent surface area considerations for the platelets and that of a spherical particle, the sheet length of the nano-platelets ($l_g$) can be converted to an equivalent theoretical diameter ($d_{eq}$) as

$$d_{eq} = \sqrt{\frac{2}{\pi}} l_g \qquad (1)$$

Although the average sheet number for the graphene sample in the present scenario is 2–5 [27], the surface area due to the platelet thickness can be neglected, since the interlayer separation in graphitic systems lies within the range of ~ 0.3–0.5 nm. Thereby, the probability of presence of the delocalized pi electron cloud at the edge of a platelet is low and since liquids exhibit phobicity towards wetting at such molecular length scales, the induced fluid polarization at the edges can be considered weak. As such, the surface area due to the platelet thickness offers very little to interfacial charge induction compared to the extended faces and can be neglected.

Functionalized graphene nanoplatelets contain chemically induced functional groups that render the intrinsically hydrophobic graphene sheets hydrophilic. The functionalization is in general performed by carboxylation or sulphonation of the graphene platelets. Spectral characterization of functionalized graphene flakes, for example, carboxylated flakes[24], reveal the presence of carboxyl (–COOH) groups as the major functional species on the surface of graphene. Consequentially, the large electronegativity of oxygen creates a massive pull on the $sp^2$ hybrid carbon oxygen bond of the carboxyl group. The partially positive carbon atom thereby poses inefficient for the $sp^3$ hybrid oxygen of the carboxyl group to attract its share of electron cloud. It therefore tends to pull off more electrons from the associated hydrogen atom, thereby creating a resonating structure where a proton and a carboxyl ion co-exist in equilibrium; while bonded to the graphene structure. The coexisting proton readily develops weak physical bonds with partly negative oxygen atoms of with the polar water molecules while the partly polarized hydrogen molecules develop hydrogen bonds with the oxygen atoms of the carboxyl group. This weak bonding in turn enhances the hydrophilicity of the graphene platelets. A possibility that the delocalized $sp^2$ $\pi$ electrons of graphene flakes also form very weak hydrogen bonds with the polar hydrogen of the water molecules is also feasible and has been illustrated in Fig. 4(b). However, owing to the low electronegativity difference between carbon and hydrogen, the bonds prove ineffective in creating long term stability of graphene in water. However, since the prediction of the charge formed at the water–platelet interface is difficult, the charge induced may be deduced from a simplistic utilization of Coulomb's law for near field electrostatics and the knowledge of the zeta potential of the colloid, as reported[28] and measured.

Conductance in colloidal media arises mainly due to two components; viz. ionic constituents and polarization of the fluid-particle interface due to formation of the Electric Double Layer (EDL). In the present case, the graphene colloids have no free ionic members and so the conductance arises solely due to polarization of the water-graphene interface. Graphene has highly delocalized pi electrons that impart high thermal and electrical conductance to it in the condensed phase. However, in an aqueous solution it proves to have a negative effect. The polar water molecules, with the oxygen atom pulling the electron cloud from the hydrogen atoms owing to its high electronegativity, develops partial negative charge on the oxygen atom. This negative partial charge imparts a repulsive effect against the delocalized electron rich graphene sheets, leading to partial hydrophobic characteristics. This phenomena of reduced wetting behavior leads to formation of a weak EDL, and thereby formation of very weak surface charge, leading to very low conductance of graphene water colloids. Functionalizing the surface enhances the EDL and the wetting characteristics by the addition of positive tail groups to the graphene surface, however, the influence of the electrons on the sheet remains strong, and low conductance is achieved.

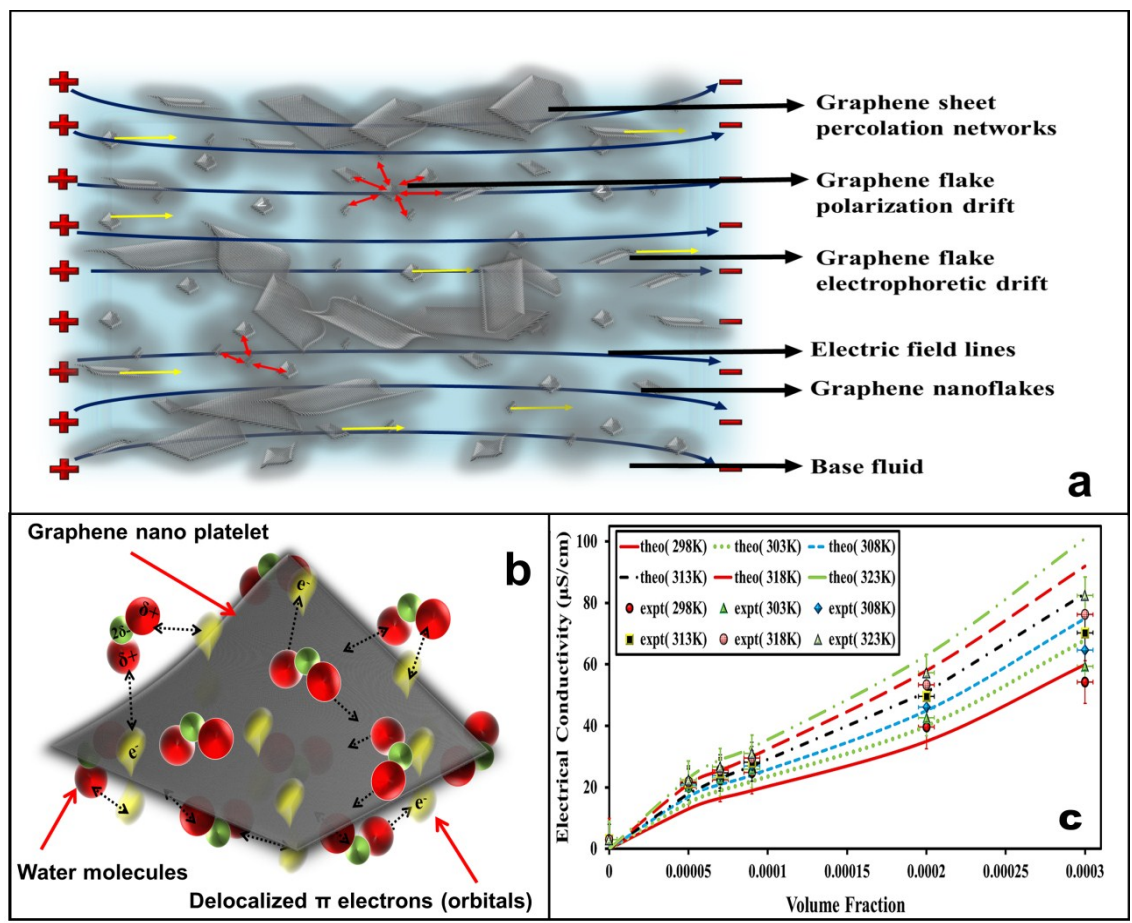

**FIGURE 4:** **(a)** A qualitative illustration of the modes of charge transport in stable graphene nano platelets-water colloids. The modes of charge transport illustrated are based on the theoretical analysis presented in an earlier work by the authors [28]. Yellow, single directional arrows indicate the direction of electrophoretic streaming, whereas, red, multidirectional group of arrows indicate the enhanced electrostatic interactions among neighboring platelets due to filed induced polarization. Blue, curved arrows indicate the imposed electric field from one electrode terminal to another. **(b)** Illustration of the mechanism in which the delocalized π electrons (shown as delocalized yellow orbital clouds) on the exposed surfaces of the functionalized graphene nanoplatelet interacts with the partially positive hydrogen atom (shown in red) polar water molecules to form the electric double layer around the platelet. The dotted arrows indicate possible weak attractions between graphene and water molecules leading to suspension stability. **(c)** Predictability of the proposed model validated against experimental observations [24]. The behavioral trends in transport with concentration and temperature have been accurately tracked by the analytical predictions. All uncertainty and error bars provided in the illustration are as reported in the concerned article.

Mathematically, the electrical conductivity of graphene nanocolloids due to electrophoretic streaming of the platelets ($\sigma_{el}$) in presence of the electric field can be expressed as[28]

$$\sigma_{el} = \frac{8\varepsilon_0^2 \kappa_f \kappa_{nc} \varsigma_{nc}^2 n \alpha \varphi_g}{d_{eq}^2 \eta_f} \qquad (2)$$

where, $\varepsilon_0$, $k_f$, $k_{nc}$, $\zeta_{nc}$, $n$, $a$, $\varphi_g$ and $\eta_f$ represent the permittivity of free space, dielectric constants of the base fluid and nanocolloid, the zeta potential of the nanocolloid, the platelet distribution parameter, the size distribution factor[23], the volume fraction of graphene and the dynamic viscosity of the base fluid respectively. Since the experimental reports lack evidences from which a probabilistic magnitude of '$a$' may be deduced, its value has been assumed to be 0.5; in accordance to findings by the present authors for chemically exfoliated graphene samples [23], wherein a similar fraction of the platelets are in the micron regime and hence non-Brownian entities. On similar grounds, the electrical conductivity of the graphene colloid due to the augmented polarization and consequential electrostatic interactions ($\sigma_p$) can be mathematically expressed as[28]

$$\sigma_p = \frac{6\varepsilon_0 \kappa_f \varsigma_{nc} \alpha n}{d_{eq}^2} \left( \frac{\kappa_g - \kappa_f}{\kappa_g + 2\kappa_f} \right) \sqrt{\frac{12\varepsilon_0 \kappa_f \varphi_g^{1/3}}{\rho_g}} \qquad (3)$$

where $\rho_g$, $k_g$ and $k_f$ represent the density of graphene and the relative permittivity of graphene and the base fluid respectively. The density and relative permittivity of graphene has been assumed equivalent to that of sub-micron graphite flakes for simplicity. The effective electrical conductivity of the graphene nanocolloid ($\sigma_{eff}$) can be predicted from the following equation[28]

$$\sigma_{eff} = \sigma_{fluid} + \sqrt{(a\sigma_{el})^2 + (b\sigma_p)^2} \qquad (4)$$

In case of nano-platelets, the coupling coefficients '$a$' and '$b$' can be intuitively theorized to possess the magnitudes of 6 and 0.16667 [28] respectively. The modified analytical model has been observed to exhibit good predictions in tune with the experimentally observed results [24]. The predictability and validity of the present model in regards to concentration of graphene and the colloid temperature has been illustrated in Fig 4(c).

**Charge Transport in G-Cl biocolloids**

The collagen intercalated graphene nano-platelets exhibit enhanced charge transport capabilities across the aqueous medium over collagen or graphene nanocolloids and with increased rate of transport at the same concentration of graphene or collagen nano platelets dispersed in the system. This by itself poses a seemingly intuitive anomaly as the transport parameters of a conglomerated system such as collagen–graphene nano–bio composites are expected to behave as a mathematical mean of the concerned parameters of the parent components. However, the expected behavioral trends are not observed, and are instead replaced by an anomalously enhanced behavior, wherein almost an order of magnitude augmentation in charge transport is observed even at dilute regimes of 0.05 wt. % graphene–collagen concentration. The experimental observations with respect to concentration and temperature of the colloidal system have been illustrated in Fig. (5).

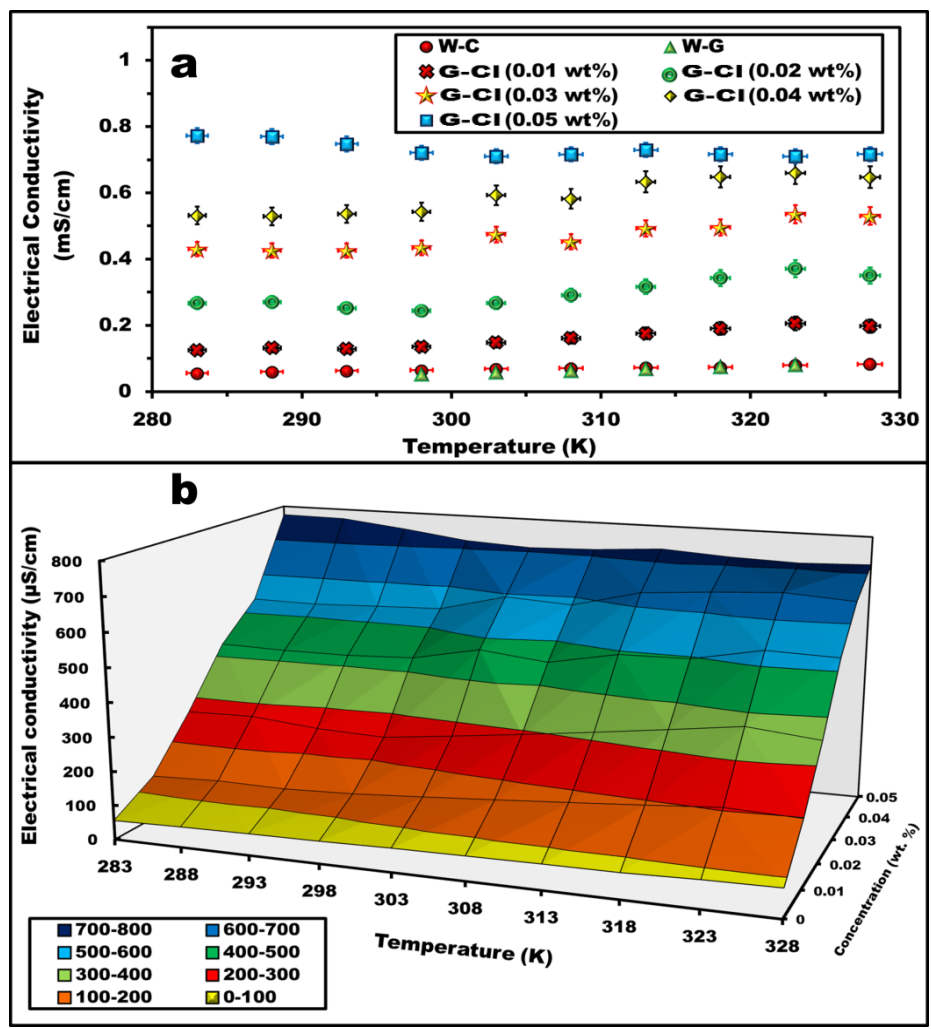

**FIGURE 5:** **(a)** Experimental observations of charge transport capabilities of G-Cl based biocolloids with respect to bio–complex concentration and temperature of the colloid. **(b)** The full spectrum of electrical conductivity of the biocolloids with respect to concentration and temperature.

The anomaly in behavior indicates an underlying mechanism via which the collagen–graphene mesh networks are directly associated with increased apparent charge of the polar colloidal system. Based on the extensive characterizations and the revelations of G-Cl interactions, it is possible to propose a mechanistic model that can predict and account for the source of such enhanced charge transport properties. The collagen fibrils, by virtue of its zwitterionic amino acid molecular entities is expected to exist as partly charged strands with both types of charges existing in unison. The majority of the charge is developed due to the proline and hydroxyproline molecules, which entwine the glycine molecules and form the external coils of the collagen fibrils. The water molecules can be theorized to interact with the collagen molecules due to the polarity on both species to form a stable collagen colloid. A weak physical bonded dispersion model can be theorized from electronegativity point of view which enables the polar water molecules to interact with the zwitterionic amino acid molecules within the collagen molecule, thus forming a stable collagen colloid. A qualitative model to represent the interaction structure and the formation of residual net charge on the unit structure has been illustrated in Fig (6).

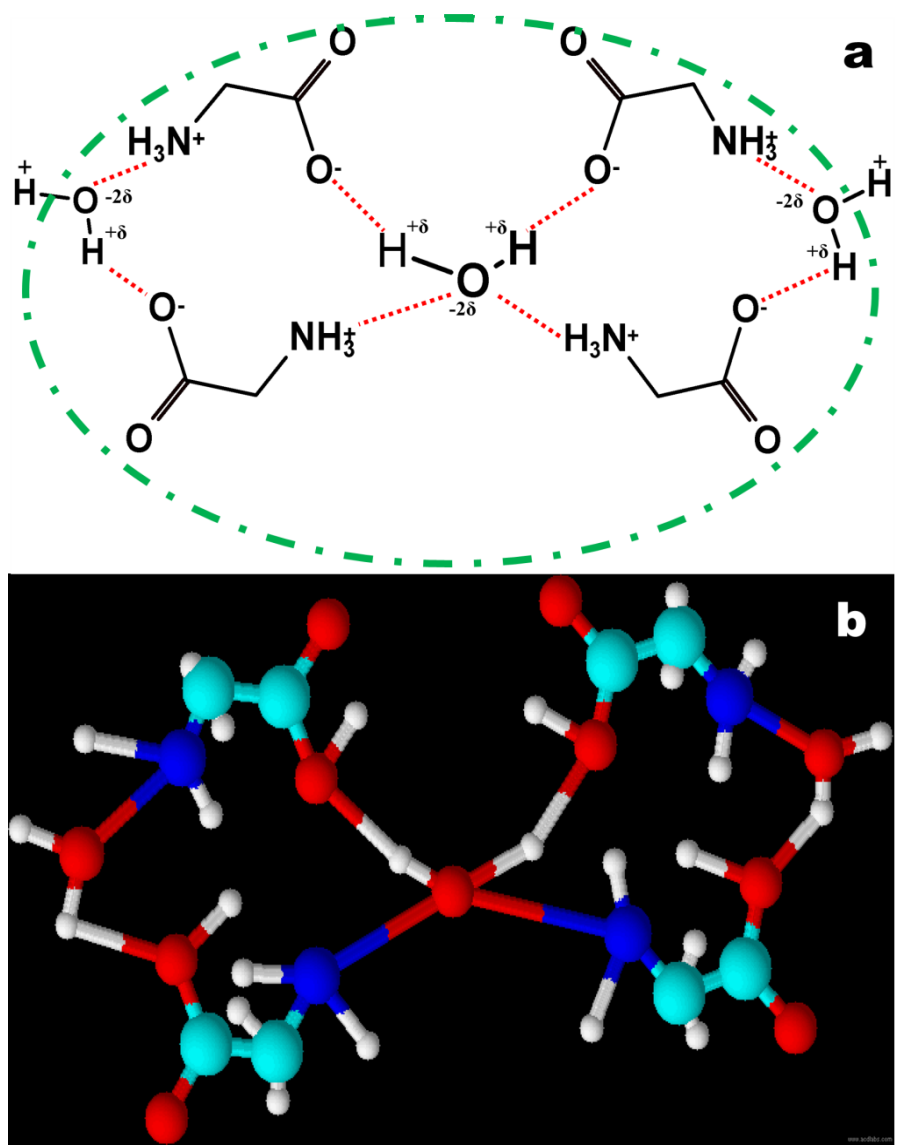

**FIGURE 6: (a)** Interaction bond diagram of collagen zwitterions with polar water molecules in the dispersed colloidal phase. Glycine molecules have been employed for reference for their structural simplicity. The dotted red lines indicate possible hydrogen bonding between two molecules owing to their electronegativity differences and presence of partial polarity. The outer green dotted boundary demarcates the feasible basic macromolecular unit formed due to interactions between water molecules and collagen fibrils, leading to partially charged structural units due the presence of twin charge unbalanced hydrogen atoms. **(b)** The 3-dimensional rendition of the base unit with only the interacting hydrogen atoms shown. The weak interactions have also been shown by solid bond lines.

Weak H-bonded structural analysis with respect to water molecules reveals that four amino acid molecules can form bonded structures with one polar water molecule. If the four molecules belong to the same collagen strand, it helps in dispersing the strand in the polar medium. In case the four molecules originate from different stands, it leads to collagen–water linkages. The structure illustrated in Fig. 6(a) is the only feasible structure involving the maximum number of collagen and water, which is expected since the system will try to reduce any charge imbalances to the farthest extent. However, it can be seen that each bonded structural unit (demarcated by the dotted frame in Fig. 6(a)) consists of charge unbalanced due to two partly polar hydrogen atoms which remain unbalanced in the structural unit. It is highly unlikely that neighboring amino acid molecules will try to neutralize the system, since they shall prefer to form their own structural units from stability point of view. Furthermore, amino acid molecules will have to be strained structurally to neutralize the charge on neighboring units. Furthermore, polar water molecules will prefer interacting with the zwitterions owing to the higher magnitude charge on the zwitterion than break hydrogen bondage to neutralize the unbalanced charge. Also, similarly charged structures ensure minimal clustering of entities, which lead to stable colloids. Thereby, each collagen fibril develops a finite residual charge when in contact with polar water molecules (a sum of all the partial unbalanced charges). The magnitude of the charge developed is dependent on the micro–fibril dimensions and collagen concentration. This partly charged structure leads to the electrical conductivity of collagen colloids. As observed from experimental findings in Fig. 5 (a), the charge developed leads to electrical transport characteristics equivalent in magnitude to graphene colloids of the same concentration.

Although dispersed, the affinity of the hydrogen molecules of water towards the functional groups on graphene leads to the delocalized planar surface $\pi$ electrons of graphene to be free from influence of the water molecules. This leads to highly dispersed platelets of graphene that are essentially charged entities, with the delocalized $\pi$ electrons acting as the source of charge. Therefore in essence, for the purpose of simplistic analysis, each such graphene platelet can be considered as a nanosheet with an effective negative charge bound to it. The previous report on collagen-intercalated graphene[27] by the present authors provides conclusive evidence (based on various characterizations) that the collagen microfibrils form physical attachments to the graphene nanoplatelets. A schematic of the collagen micro–fibrillated graphene nanoplatelet system has been illustrated in Fig. 7. In aqueous collagen colloids, the partly positive hydrogen atom of the amine groups and the partly negative oxygen of the carboxyl groups of the amino acid molecules are expected to form weak hydrogen bonds with the polar water molecules, thus forming a stable colloid. However, owing to the high electron density on the graphene flakes (due to the highly delocalized π electron cloud which provide graphene a semi-metallic character) when compared to a single water molecule, the partly positive amine group now loses its affinity for the oxygen of the neighboring water molecules and forms weak bonds with the graphene platelet. This is also possible due to the intercalation of the collagen into graphene flakes as reported in previous report. The high electron density of the graphene platelet leads to the formation of a bond of strength high enough to resemble a chemical bond, as evident from characterizations[27]. Based on physical bonding, a structure can be theorized, wherein the polar water molecules, the zwitterionic amino acid molecules and the charged graphene platelets form a mesh or network. The corresponding illustration of the same has been provided in Fig. 7.

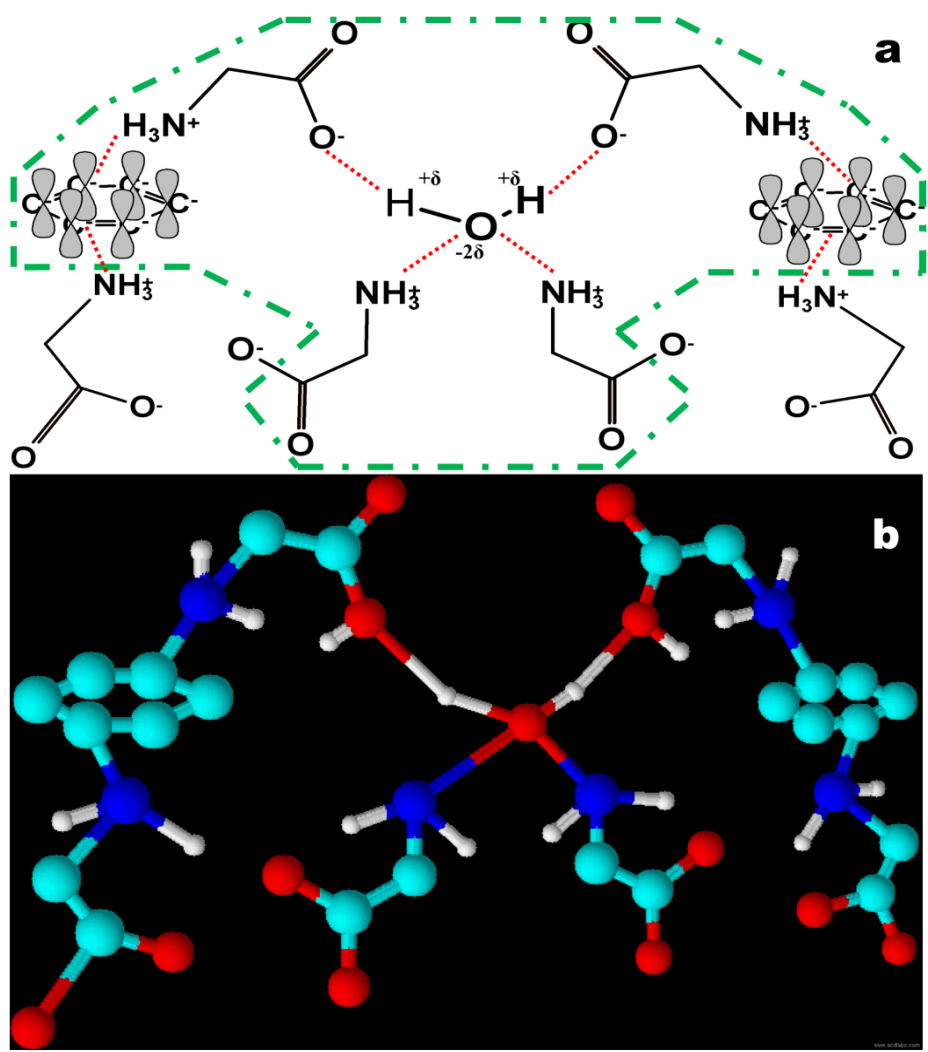

**FIGURE 7**: The interaction of graphene-collagen intercalated bio–complex with polar water molecules **(a)** molecular line schematic of bond formation of collagen zwitterion molecules with graphene. The higher electron density in graphene causes the positive amine group to bond with the flakes. The negative carboxyl group thereby remains with the only choice of grouping with a neighboring water molecule. Based on charge minimization, the present structure is the most feasible. **(b)** The 3 dimensional diagram of the graphene–collagen system within a polar dispersion phase.

The enhanced charge transport is observed solely in the aqueous colloidal state. The enhanced transport arises due to enhanced charge within the electric double layer (EDL), which is a result of the interactions of the zwitterionic amino acid molecules in the collagen micro-strands. Existence of the zwitterions is only possible when the stands are dispersed in a highly polar solvent, in this case water. Therefore, hydration is the key to augmented transport. The transport is not electronic, but ionic since the augmentation is due to formation of enhanced surface charge, leading to enhanced polarity of the individual graphene flakes. If de-hydrated, the graphene-collagen complex still remains in the physically bonded structure; however, the formation of the surface charge will cease due to the absence of zwitterions. The charge transport studies on purely graphene colloids and collagen colloids act as the controls. The surface charge formation for these colloids can be assessed and based on that, the interfacial charge development on the graphene-collagen complex can be determined. Enhanced protonation of the colloid by addition of weak acids leads to enhanced stability due to shifting of the zeta potential away from the iso–electric point. However, charge transport due to the graphene-collagen complex solely cannot be comprehended from such a system as the protons will also contribute to the transport process.

It is evident from the mesh structure that the structural unit in the case of the bio-composite is grossly different to that of the collagen colloids. It is noteworthy that the structure has been theorized on the grounds of minimal unbalanced charge on the unit domain, and thereby is the most feasible. Interestingly, however, even in this case the bond link structure predicts that the structure is based around a single polar water molecule as in the case of collagen. However, presently, the unbalanced charges arise from the amino acid molecules of the collagen strands, and not from the partly polar water molecules. It can be observed that the net charge in the present structural unit is twice that of a single collagen colloid structural unit, but opposite in charge. Moreover, since the graphene platelets act as an intermediate entity with dense delocalized electron cloud, there can be multiple collagen microfibrils attached to the same graphene platelet. This has also been established in the previous report[27]. Thereby, the graphene–collagen complex based biocolloids transport charge via three mechanisms, viz.; transport by the graphene flakes similar to basic graphene colloids (since they are still free to move by electrophoresis, owing to the length and soft fibril nature of collagen stands), transport by the partly polarized water molecules in interaction with collagen zwitterions and finally due to the formation of excess partly polar collagen molecules due to interactions by graphene–collagen complex with water molecules. Thereby, in essence, the transport capability increases by a factor of nearly four folds than the base graphene or collagen colloids. Furthermore, the point of interest lies in the fact that Dynamic Light Scattering analyses[27] exhibits formation of graphene flakes in the nano and submicron range due to biomimetic exfoliation by collagen. Thereby, the graphene differs from the conventional chemically synthesized graphene with respect to the platelet size and nearly the whole population consists of Brownian platelets. Consequently, the magnitude of '$a$' also tends towards unity, i.e. nearly doubles itself. As a result, at similar concentrations, almost an eight fold increase in electrical conductivity is expected. This is verified from the experimental data (Fig. 4) where the conductivity of the 0.05 wt. % graphene-collagen biocolloid increases nearly an order of magnitude than the base graphene or collagen colloids of same concentrations. Since the experimental value of '$a$' has been assumed from similar findings and the mechanistic theory proposed is phenomenological by nature, the deviations of the theoretical predictions from the experimental data are expected. Also, as expected from Eqn. (2) and (3), the degree of augmentation reduces almost linearly with decreasing concentration and similar trends are observed in experimental data. Interestingly, the fact that the mechanistic theory can predict the increase in transport parameters accurately to within ±20 % of the observed values speak of the validity of the mechanisms and physics proposed.

## Conclusion

To infer, graphitic structures have been bio-mimetically exfoliated by Cl microfibrils to form graphene-collagen bio composites, which have been subjected to extensive characterizations to understand the mechanism of intercalation. The electrical transport capabilities of synthesized[27] graphene–collagen biocomposite based aqueous biocolloids have been studied. The colloids have been observed to yield anomalously higher electrical conductivities than the base graphene or collagen based colloids. The anomaly lies in the fact that graphene and collagen colloids, both of low electrical conductivities, yield high conducting colloid when the biocomposite is utilized. An analytical model to explain the physics behind transport of charge in graphene colloids and to mathematically deduce the electrical conductivity from system properties has been proposed and found to behave in consistence the experimental observations reported in literature. The G-Cl composites have been subjected to extensive characterizations so as to understand the interaction mechanisms of Cl with graphenic structures. Based on such comprehensive details, a mechanism to explain the source of electrical conductance of collagen colloids has also been developed and put forward. On similar grounds, a mechanistic approach to evolve a phenomenological reasoning so as to comprehend the mechanism and predict the augmented transport in the complex biocolloids has been theorized. Based on experimental evidences, a physico–chemical approach to understand the interactions between graphene, zwitterionic amino acids in collagen and polar water molecules has been theorized. Based on the proposed theory, the augmented transport capability can be predicted within ±20% of the experimental data from simplistic logical analysis. The accuracy provides support to the strength of the theory and further heightens the promise of the novel graphene–collagen biocolloids as potential candidates for advanced drug delivery, DNA gel electrophoresis medium, active transport bio-colloids and other potential and/or future medical applications.

## Acknowledgements

P.D would like to acknowledge the Ministry of Human Resource and Development, Government of India for the doctoral research scholarship.